\documentclass[osajnl,twocolumn]{revtex4}
\usepackage{graphicx}
\usepackage{hyperref}

\begin{document}

\title{Cluster state quantum computing in optical fibers}
\author{Yasaman Soudagar}
\email{yasaman.soudagar@polymtl.ca}
\author{F\'elix Bussi\`eres}
\altaffiliation[Also at ]{Laboratoire d'informatique th\'eorique et quantique,
Universit\'e de Montr\'eal, C.P. 6128, Succ. Centre-Ville, Montr\'eal (QC),
H3C~3J7 Canada}
\author{Guido Berl\'\i n} 
\author{Suzanne Lacroix} 
\author{Jos\'e M. Fernandez}
\altaffiliation{ \'Ecole Polytechnique de
Montr\'eal, Department of Computer Engineering}
\author{Nicolas Godbout} 
\affiliation{Laboratoire des fibres optiques,
Centre d'optique, photonique et laser\\ \'Ecole Polytechnique de
Montr\'eal, Engineering Physics Department\\ P.O. Box 6079, Station
Centre-ville, Montr\'eal (Qu\'ebec) H3C~3A7 Canada}

\begin{abstract}
A scheme for the implementation of the cluster state model of quantum computing in optical fibers, which enables the feedforward feature, is proposed. This scheme uses the time-bin encoding of qubits. Following previously suggested methods of applying arbitrary one-qubit gates in optical fibers, two different ways for the realization of fusion gate types~I and II for cluster production are proposed: a fully time-bin based encoding scheme and a combination of time-bin and polarization based encoding scheme. Also the methods of measurement in any desired bases for the purpose of the processing of cluster state computing for both these encodings are explained.
\end{abstract}
\ocis{Fiber optics and optical communications (060.0060), Quantum optics (270.0270)}

\maketitle

\section{Introduction}

\begin{figure}[!h]
\centerline{\includegraphics[scale=1]{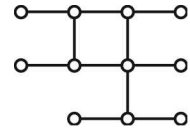}}
\caption{An arbitrary cluster: Each of the vertices represents a qubit in the $|+\rangle$ state. A controlled-Z gate\cite{CZ} is applied between each pair of the qubits which are connected by an edge. The controlled-Z operations can be applied in any order, since they commute when applied on $|+\rangle$ states.}
\label{fig:graph}
\end{figure}

One of the major difficulties in the implementation of quantum computers using optical systems is the weakness of nonlinear interactions between photons.  This makes the implementation of two-qubit gates, where each qubit is encoded on a photon, a daunting task.  An implementation of probabilistic multi-qubit gates using only linear optical elements was proposed,\cite{KLM} but it requires very large optical circuits with interferometric phase stability in every path.  The cluster state model of quantum computing\cite{Raussendorf and Briegel} overcomes these difficulties by moving the complexity to the generation of large clusters of entangled states and by reducing the task of processing to 1-qubit measurements in certain bases.  In this model, a cluster of entangled qubits can be represented by a two dimensional graph, such as in Fig.~\ref{fig:graph}, where each vertex represents a qubit prepared in state $|+\rangle = (|0\rangle + |1\rangle)/\sqrt 2$ and each edge signifies that a controlled-Z gate\cite{CZ} was applied between connected vertices. After creation of such a cluster, the processing, that includes only single qubit measurements in certain bases, begins. The qubits to be measured, the order of measurement and the measurement bases are determined by the algorithm.  The first set of qubits are measured each in a pre-defined basis. The results of these measurements are fedforward to set the bases for the second set of measurements. This process continues until the completion of the algorithm. The remaining unmeasured qubits, which contain the output, might require some single qubit corrections. In this model, the information is transferred from one set of measured qubits to the next, essentially through 1-qubit teleportation,\cite{Nielsen} and the processing of information is applied through the choice of bases for measurements.  Experimentally, the most difficult part in the optical realization of this model is the production of the initial cluster.

\begin{figure}
\centerline{\includegraphics[scale=1.05]{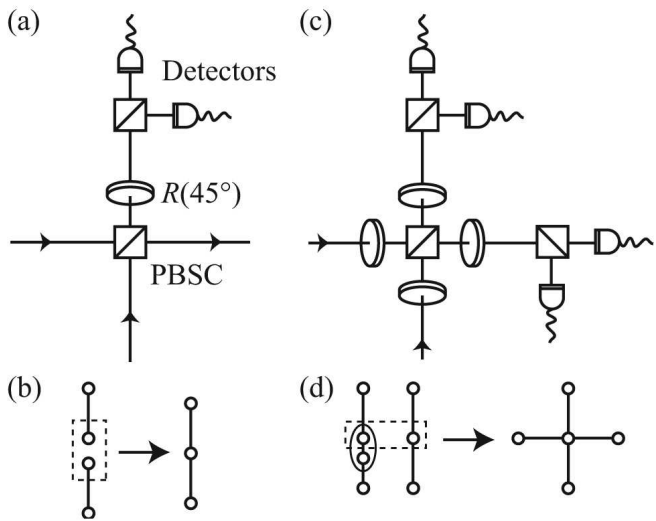}}
\caption{Fusion gates as proposed by Brown and Rudolph:~\cite{Browne and Rudolph} PBSC is polarization beam splitter/combiner and the rotation gate $R(45^\circ)$ is applied using a half-wave plate. In these gates, the qubits are encoded on the polarization of photons, such that $\vert 0 \rangle$ corresponds to a horizontally polarized photon, denoted by $\vert H \rangle$ and $\vert 1 \rangle$ corresponds to a vertically polarized photon, denoted by $\vert V \rangle$. One photon (enclosed in dashed boxes) from each cluster to be fused is sent through the fusion gate such that each photon enters the gate on a different spatial path.  (a)~Fusion gate type~I: The fusion is successful if only one of the detectors detects exactly one photon, which happens with probability $1/2$.  (b)~In case of success, two clusters are fused together to create a larger one.  (c)~Fusion gate type~II: The fusion is successful if one of the detectors fires after each PBSC.  The probability of success is $1/2$. (d)~The two qubits enclosed in the oval shaped solid line are redundantly encoded.  The resulting cluster, in case of sucess, is shown.}
\label{fig:fusiongates}
\end{figure}

It has been shown that using non-deterministic \emph{fusion gates},\cite{Browne and Rudolph} it is possible to produce clusters with a reasonable amount of resources. These gates ``fuse'' two clusters together to create a larger one, as shown in Fig.~\ref{fig:fusiongates} for polarization encoding. In order to do so, according to the desired shape of the resulting cluster, one chooses a qubit from each of the clusters which are to be fused.  These qubits go through the fusion gate and if the gate is successful, which happens with probability $1/2$, the two clusters will be fused together. In fusion gate type~I (Fig.~\ref{fig:fusiongates}a and~\ref{fig:fusiongates}b), the fusion is successful if only one of the detectors detects exactly one photon.  The effect of failure is equivalent to the measurement of these two photons in a basis which breaks the cluster at the measured vertex into two smaller clusters.  In fusion gate type~II (Fig.~\ref{fig:fusiongates}c and~\ref{fig:fusiongates}d), the fusion is successful if after each polarization beam splitter, one of the detectors fires.  The effect of failure is equivalent to the measurement of photons in a basis such that it creates a rather useful redundant encoding, keeping the entanglement of the two clusters intact.  In fact, a good way to create a cluster of the desired shape with type~II fusion is to use redundant encoding, as shown in Fig.~\ref{fig:fusiongates}d, which can be created by measurement in a specified basis.  Two-qubit clusters can be used as the seed to create larger ones. They can be efficiently created using parametric down conversion and the application of some single qubit gates.

So far, a number of very interesting experiments towards the implementation of the cluster state model and fusion gate type~I have been performed.\cite{ZeilingerNature,Pan Fusion I} These experiments were done in free space and used photon polarization as the physical encoding of qubits.  Although this encoding is very advantageous due to the ease of polarization manipulation in free space, implementing the feedforward of measurement results, which is an essential feature of the cluster state model, is not obvious. Allowing for a 10~nanosecond delay for one measurement, plus classical processing and feedforward to an active optical device, the remaining photons of the cluster must be stored in an optical delay line of at least three-meter optical path at each processing cycle.  The propagation of beams in free space over distances of tens of meters suffers from two hurdles to achieve exact mode overlap necessary for the multi-path interferometric measurements. 1)~Diffraction: a gaussian beam of typical waist size of 1~mm suffers from noticeable diffraction after propagation over its Rayleigh length, which is 2~meters. 2)~Decreasing tolerance to pointing errors: a 100~$\mu$m overlap accuracy of two beams after 30~meters of propagation requires a source with pointing accuracy of 3~microradians.  These two hurdles are eliminated in optical fibers since the mode profile is conserved over arbitrary large propagation distances.  Optical fiber is therefore an ideal candidate for storing photons with high fidelity over these distances.

\begin{figure}
\centerline{\includegraphics[scale=1]{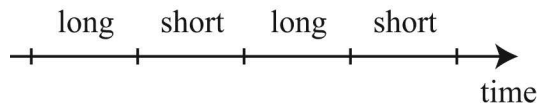}}
\caption{In our time-bin encoding we divide the time of arrival of photons at the detector to alternate between states $\vert s \rangle$ and $\vert l \rangle$.}
\label{fig:time-bin}
\end{figure}

In this article, we propose the realization of quantum computing using the cluster state model in optical fibers. In our proposed implementation, each vertex corresponds to a single photon. Each of these photons is transmitted in its own optical fiber.  The qubits are encoded using time-bin encoding.\cite{Gisin99} In this encoding, we divide the time line of the arrival of photons at detectors into a sequence of time slots, or bins, with alternating logical values (Fig.~\ref{fig:time-bin}).  The bins are labeled $|0\rangle$ = $|s\rangle$ and $|1\rangle$ = $|l\rangle$, where $s= \mathrm{short}$ and $l = \mathrm{long}$. One can create an arbitrary $\alpha|s\rangle + \beta|l\rangle$ state by splitting a single-photon light pulse through a delay-line interferometer with adjustable splitting ratio and phase between the short and long delay lines.  A 2-qubit cluster can be generated by preparing a high-power optical light pulse in the state $(|s\rangle+|l\rangle)/\sqrt 2$ and sending it through a parametric down-converter, resulting in an entangled photon pair in the state $(|ss\rangle + |ll\rangle)/\sqrt 2$.  With the application of an adapted Hadamard gate (discussed in section~\ref{sec:proctimebin}) to the second qubit, this state will change into $(|ss\rangle +|sl\rangle + |ls\rangle - |ll\rangle )/ 2$, which is exactly the state for a 2-qubit cluster. Several such 2-qubit clusters are used as seeds to be fused together to create larger clusters.

Time-bin entanglement is experimentally proven to be robust against decoherence after transmission over more than 25~km of fiber,\cite{Gisin04} which corresponds to a propagation time of the order of $10^{-4}$ seconds. On the other hand, it is shown that the lifetime of entanglement under decoherence for the cluster type entanglement does not depend on the size of the system.\cite{Dur and Briegel} This shows that, in principle, it is possible for a cluster of entangled photons to keep its full entanglement in optical fibers long enough to allow the completion of the computation, including the feedforward of measurement results and setting the new measurement bases accordingly. By using electronic devices with a 100~MHz bandwidth, these operations can be performed in the order of tens of nanoseconds.

This article is divided as follows. Fiber optical circuits for the production of cluster states and their processing in the fully time-bin encoding scheme are presented.  An alternative scheme, using a combination of time-bin and polarization encoding, is also presented.  We finally briefly discuss experimental aspects of the proposed schemes.

\section{Cluster production and processing}

\subsection{Cluster production using the fully time-bin based scheme}

The proposed fusion gates in Ref.~\onlinecite{Browne and Rudolph} consist of a polarization beam splitter/combiner (PBSC) and $45^\circ$ polarization rotators,  denoted $R(45^\circ)$ (Fig.~\ref{fig:fusiongates}). A polarization beam splitter/combiner allows the horizontally polarized light to go through, and it reflects vertically polarized light. The equivalent of the PBSC for time-bin encoded qubits is an active, e.g. electrooptic, switch. The switch has two inputs and can be set, at any given time, such that either both inputs continue on their original rail or they switch rails. Hence one can set it to allow qubits in the time slots corresponding to the short bin ($|s \rangle$) to continue on their original rail, and those in the long time-bin ($\vert l \rangle$) to switch rails.

\begin{figure}
\centerline{\includegraphics[scale=1]{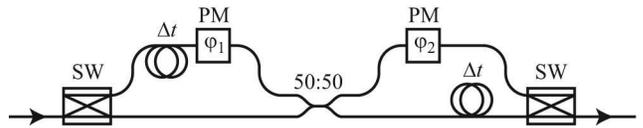}}
\caption{Reconfigurable time-bin one-qubit gate. SW: active switch; PM: phase modulator. This gate applies an $R(45^{\circ})$
gate to single qubits when both $\varphi_1$ and $\varphi_2$ are
set to $\pi$. It applies a Hadamard gate by setting $\varphi_1$ to
zero and $\varphi_2$ to $\pi$.}
\label{fig:45Rotate}
\end{figure}

Although $R(45^\circ)$ is very easy to implement for polarization encoding, its counterpart for time-bin encoding is not as simple. The $R(45^\circ)$ gate applies the transformation
$$ \vert 0 \rangle \rightarrow \frac{\vert 0 \rangle+\vert 1 \rangle}{\sqrt
2},\;\;\;\;\; \vert 1 \rangle \rightarrow \frac{-\vert 0 \rangle+\vert 1
\rangle}{\sqrt 2}$$ 
therefore
$$ \frac{\vert 0 \rangle+\vert 1 \rangle}{\sqrt 2} \rightarrow \vert 1
\rangle, \;\;\;\;\; \frac{\vert 0 \rangle-\vert 1 \rangle}{\sqrt2}
\rightarrow \vert 0 \rangle.$$ 
To realize this transformation in a time-bin encoding, one adapts the universal gate for a fully time-bin scheme discussed in Ref.~\onlinecite{OURS} as shown in Fig.~\ref{fig:45Rotate}.  We denote this adapted gate for time-bin as $R_t(45^\circ)$. To see how this gate works, one sends a photon in state $\alpha \vert s \rangle + \beta \vert l \rangle $ to the input rail of the gate.  The first optical switch sends the state $\vert s \rangle$ to the upper (long) rail and the state $\vert l \rangle$ to the lower (short) one.  The two states are synchronized before the 50:50 beamsplitter,\cite{BeamSplitter} while the one on the upper rail acquires an additional phase.  The two states interfere at the beamsplitter. Then the one on the upper rail experiences a phase shift and that on the lower rail experiences a delay to put the states back to different time-bins, before they are all transfered to the lower rail by the last switch.  For our specific input, the output of the circuit is $(\alpha - \beta)/\sqrt 2 \vert s \rangle + (\alpha +\beta)/\sqrt 2 \vert l \rangle $, which is the desired result.

\begin{figure}[t]
\centerline{\includegraphics[scale=1]{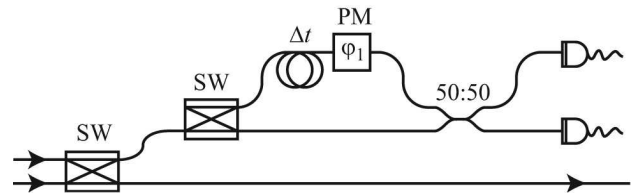}}
\caption{Fusion gate type~I for time-bin encoded qubits. Here $\varphi_1 = \pi$ and fusion is successful if only one of the detectors detects exactly one photon.}
\label{fig:typeI}
\end{figure}

Figure~\ref{fig:typeI} shows the complete fusion gate type~I for time-bin encoding. One can show that the last switch and detector of the $R_t(45^\circ)$ gate can be substituted by two detectors on each rail.\cite{OURS} This fusion gate succeeds if exactly one photon is detected. The probability of success of the gate is $1/2$.

\begin{figure}[t]
\centerline{\includegraphics[scale=1]{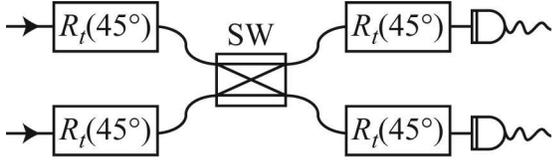}}
\caption{Fusion gate type~II for time-bin encoded qubits. $R_t(45^\circ)$ is the gate depicted in Fig.~\ref{fig:45Rotate}. Fusion is successful if both detectors fire.}
\label{fig:typeII}
\end{figure}

The reconfigurable single qubit gate discussed above can be used for the implementation of a fusion gate type~II as shown in Fig.~\ref{fig:typeII}.  Four $R_t(45^\circ)$ gates are used before and after the optical switch that acts as the PBSC. The gate is successful when both detectors fire and, as mentioned in Ref.~\onlinecite{Browne and Rudolph}, there is no need for photon counting detectors.  This gate is successful with probability $1/2$.

\subsection{Processing in the fully time-bin based scheme}
\label{sec:proctimebin}

\begin{figure}
\centerline{\includegraphics[scale=1]{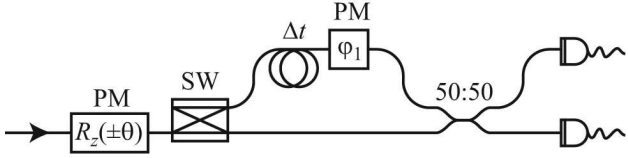}}
\caption{Detection in a desired basis using the fully time-bin scheme. The variable phase modulator before the first switch applies $R_z(\pm\theta)$, where the sign is determined according to the results of the previous set of measurements.  The rest of the circuit enables measurement in the Hadamard basis and is based on the reconfigurable gate of Fig.~\ref{fig:45Rotate} with $\varphi_1 = 0$.}
\label{fig:timebinproc}
\end{figure}

In order to create the 2-qubit clusters to be used as seeds for the creation of larger clusters and also to perform measurements in the required basis, the Hadamard gate is necessary. This gate transforms
$$ \vert 0 \rangle \rightarrow \frac{\vert 0 \rangle+\vert 1
\rangle}{\sqrt 2}, \;\;\;\;\; \vert 1 \rangle \rightarrow \frac{\vert
0 \rangle-\vert 1 \rangle}{\sqrt 2}$$
therefore
$$\frac{\vert 0 \rangle+\vert 1 \rangle}{\sqrt 2} \rightarrow \vert
0
 \rangle, \;\;\;\;\; \frac{\vert 0 \rangle-\vert 1
\rangle}{\sqrt2}
 \rightarrow \vert 1 \rangle.$$
To obtain the Hadamard gate in time-bin encoding, one can use the same gate shown in Fig.~\ref{fig:45Rotate}, with $\varphi_1 = 0$ and $\varphi_2 = \pi$.   After the creation of the cluster of entangled photons is completed, one begins the processing, which consists of single qubit measurements in different bases. Performing these measurements is equivalent to applying an $R_z(\pm\theta)$ followed by a Hadamard gate to the qubit before detection in the computational basis.\cite{Rz} For time-bin encoding, $R_z(\pm\theta)$ is performed with a time-varying phase modulator, which applies a phase $-\theta/2$ to state $|s \rangle$ and phase $\theta/2$ to state $|l \rangle$.  Figure~\ref{fig:timebinproc} depicts the measurement scheme. 

The algorithm, which is being implemented using the cluster state model of computation, determines the absolute value of the argument of $R_z(\pm\theta)$. However, the sign of the argument, at each step of the computation, depends on the results of the previous set of measurements. One has to wait for these results to be fedforward to the phase modulator to set the correct value of the angle $\pm\theta$.  While this is happening, the cluster of time-bin entangled photons is kept in optical fiber loops.

\begin{figure}
\centerline{\includegraphics[scale=1]{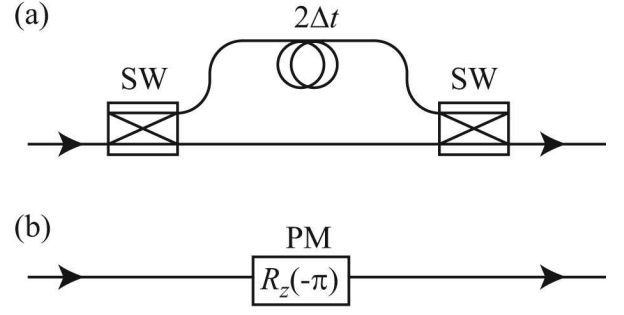}}
\caption{Circuits suitable for the 1-qubit corrections to the remaining qubits of the cluster.  (a)~Bit flip operation for time-bin qubits.  The $|s\rangle$ state is delayed by $2\Delta t$, i.e. twice the delay between short and long bins. (b)~Phase flip operation is equivalent to $R_z(-\pi)$ up to a global phase. It applies a $\pi$ phase shift difference to $|s\rangle$ and $|l\rangle$.}
\label{fig:corrections}
\end{figure}

After all the processing is done, depending on the measurement results, one might need to apply single qubit corrections to the output. These corrections are only of two different kinds. One is a bit flip, which changes $|s\rangle$ to $|l\rangle$ and $|l\rangle$ to $|s\rangle$ and can be achieved using the circuit shown in Fig.~\ref{fig:corrections}a.  The second one is a phase flip, which changes the $|1\rangle$ state to $-|1\rangle$ and leaves the state $|0\rangle$ unchanged.  This can be achieved with a phase modulator as shown in Fig.~\ref{fig:corrections}b.

\subsection{Cluster production using the polarization based scheme}

\begin{figure}
\centerline{\includegraphics[scale=1]{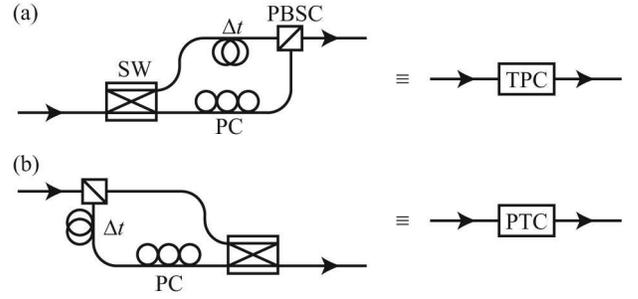}}
\caption{Encoding convertors: PC is a polarization controller and PBSC is a polarization beam splitter/combiner.  The labels TPC and PTC stand for time-bin to polarization convertor and polarization to time-bin convertor. (a) This setup converts time-bin encoding to polarization encoding. One can set the switch and the polarization controller such that state $|s\rangle$ is converted to $|H\rangle$ and state $|l\rangle$ to $|V\rangle$. (b) Using this setup, polarization encoding is converted to time-bin encoding.}
\label{fig:convert}
\end{figure}

\begin{figure}
\centerline{\includegraphics[scale=1]{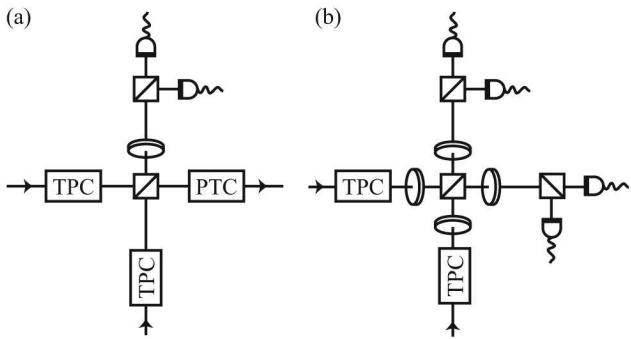}}
\caption{(a) Fusion gate type~I.  Time-bin encoded qubits are converted to polarization encoding before they go through the gate. The output is converted back to time-bin encoding. The gate is successful, with probability $1/2$, when one of the detectors detects one photon. (b) Fusion gate type~II: This gate is successful, with probability $1/2$, when one detector fires after each PBSC.}
\label{fig:FusionPol}
\end{figure}

Another way to implement fusion gates for time-bin encoding is to use the method described in Ref.~\onlinecite{OURS} to convert the states of each of the photons from time-bin encoding to polarization encoding. In other words, one converts the state $\vert s \rangle$ to $\vert H \rangle$ and $\vert l \rangle$ to $\vert V \rangle$ with the circuit depicted in Fig.~\ref{fig:convert}a.  One can then use the exact techniques discussed in Ref.~\onlinecite{Browne and Rudolph} for polarization encoded qubits to apply the fusion gates. The fusion gate type~I depicted in Fig.~\ref{fig:FusionPol}a requires a polarization beam splitter/combiner. A polarization controller is placed after one output to rotate the polarization by $45^\circ$ before the polarization discriminating photon counting detection.  The other output goes through the fiber optical circuit that converts the polarization state back to time-bin, namely $\vert H \rangle$ to $\vert s \rangle$ and $\vert V \rangle$ to $\vert l \rangle$, depicted in Fig.~\ref{fig:convert}b.

Fusion gate type~II is similar to type~I, except that it requires three more polarization controllers. Also one does not require to convert the qubits back to time-bin encoding before detection. The setup is shown in Fig.~\ref{fig:FusionPol}b.

\subsection{Processing in the polarization based scheme}

\begin{figure}
\centerline{\includegraphics[scale=1]{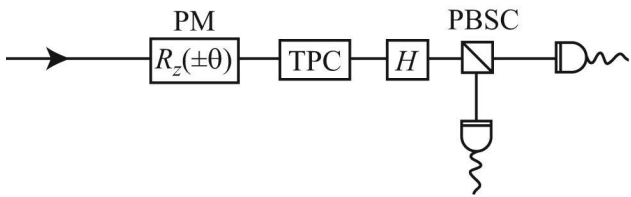}}
\caption{Detection in a desired basis using the polarization based scheme.  A Hadamard gate is applied in polarization encoding by using a half-wave plate.}
\label{fig:PolarizationProcess}
\end{figure}

In the processing stage, to carry on the measurement in any desired bases, one converts the time-bin qubits to be measured to polarization qubits and, using the usual devices for polarization manipulation, one applies the required rotation before detection. Figure~\ref{fig:PolarizationProcess} shows the proper setup to achieve this goal. The eventual single qubit corrections to the output, bit-flip and phase-flip, can be applied using wave-plates.

\section{Discussion}

The proposed experimental setups can be implemented using currently available optical technology.  All-fiber components, such as couplers (the all-fiber equivalent of beamsplitters) and polarization controllers have very low insertion loss.  Fiber-pigtailed bulk components, such as polarization beamsplitter/combiners, also have low optical loss.  Active electrooptic components, such as the switches and phase modulators, are readily available with high bandwidth of 10~GHz or more, allowing for sub-100~ps switching.  However, such active components currently impose significant excess loss of the order of 30--60\%.  The amount of photon loss through several electrooptic devices limits the proposed schemes to small clusters.  However, it is worth mentioning that the optical losses of current electrooptic components constitute a purely technological problem which can be expected to be mitigated over the next few years, as typically happens with standard optical telecommunication devices.  A thorough analysis of the experimental scaling of computing efficiency with the number of qubits is warranted, taking into account optical losses, detector efficiency, initial photon statistics and other deleterious effects.  This scaling analysis will be presented later.

\section{Conclusion}
In this article, we have proposed the realization of the cluster state model of quantum computing in optical fibers, which enables the inclusion of the feedforward of the measurement results, an essential feature of this model.  We have proposed two different experimental schemes leading towards the realization of fusion gates type~I and type~II, using either time-bin encoding or a combination of time-bin and polarization encoding.  The proposed schemes can be implemented for small clusters using currently available technology.  Their use for larger clusters requires improved active devices such as switches and phase modulators with lower optical losses, which can be reasonably expected within a few years.  Future work includes the experimental demonstration of the proposed schemes for small clusters and an analysis of their scaling to the large ones.

\section*{Acknowledgments}

Y. S. wishes to thank Dr. Elham Kashefi for useful discussion on the Cluster State model.  This work is supported by the Canadian Institute for Photonics Innovations (CIPI) and by the Natural Sciences and Engineering Research Council of Canada (NSERC).


\begin{thebibliography}{99}
\bibitem{CZ}
The controlled-Z gate is represented by the matrix
\begin{displaymath}
\left( 
\begin{array}{cccc}{1} & {0} & {0} & {0} \\
{0} & {1} & {0} & {0}\\
{0} & {0} & {1} & {0}\\
{0} & {0} & {0} & {-1}\\
\end{array}
\right ).
\end{displaymath}
It changes the state $+|11\rangle$ to $-|11\rangle$ and leaves the
states $|00\rangle$, $|01\rangle$ and $|10\rangle$ unchanged.
\bibitem{Rz}
$R_z(\theta)$ is represented by matrix 
\begin{displaymath}
\left(
\begin{array}{cc}{e^{-i\theta/2}} & {0} \\
{0} & {e^{i\theta/2}}
\end{array}
\right ),
\end{displaymath}
which is equivalent to applying the phase $-\theta/2$ to state $| 0 \rangle$ and phase $+\theta/2$ to state $|1 \rangle$.
\end{thebibliography}
\end{document}